\documentclass[twocolumn]{elsart}
\usepackage{graphicx}
\usepackage{dcolumn}
\begin{document}
\runauthor{Calloni}
\begin{frontmatter}
\title{The Aladin2 experiment: sensitivity study}
\author[FedII and INFN]{G. Bimonte \thanksref{PRIN}}
\author[Sec and INFN]{D. Born}
\author[FedII and INFN]{E. Calloni}
\author[FedII and INFN]{G. Esposito \thanksref{PRIN}}
\author[IPHT]{U. H$\ddot{u}$bner}
\author[IPHT]{E. Il'ichev}
\author[FedII and INFN]{L. Rosa}
\author[FedII and INFN]{O. Scaldaferri}
\author[Sec and INFN]{F. Tafuri}
\author[FedII and INFN]{R. Vaglio}
\address[FedII and INFN]{INFN, Sezione di Napoli, and Dipartimento 
di Scienze Fisiche, Complesso Universitario di Monte S. Angelo, 
Via Cintia, Edificio N', 80126 Napoli, Italy}
\address[Sec and INFN]{Seconda $Universit\grave{a}$ di Napoli and INFN, 
Sezione di Napoli,  Complesso Universitario di Monte S. Angelo, 
Via Cintia, Edificio N', 80126 Napoli, Italy}
\address[IPHT]{Institut F$\ddot{u}$r Physikalische HochTechnologie e. V., 
Postfach 10 02 39 - 07702 Jena, Germany}
\thanks[PRIN]{Partially supported by PRIN {\it SINTESI}}
\begin{abstract}
Aladin2 is an experiment devoted to the first measurement of 
variations of Casimir energy in a rigid body. The main short term 
scientific motivation relies on the possibility of the first 
demonstration of a phase transition influenced by vacuum fluctuations 
while, in the long-term and in the mainframe of the cosmological 
constant problem, it can be regarded as the first step towards a  
measurement of the weight of vacuum energy. In this paper, after 
a presentation of the guiding principle of the measurement, 
the experimental apparatus and sensitivity studies on final cavities 
will be presented.
\end{abstract}
\end{frontmatter}

\section{Introduction}

Although phase transitions influenced by vacuum fluctuations are 
theoretically predicted to play a fundamental role in cosmology 
there is not, at present, an experimental demonstration of such a
transition, not even at the microscopic scale.
Furthermore, even if much progress has been made in the evaluation 
and experimental verification of effects produced by vacuum energy in
Minkowski space-time \cite{Moste}, it remains unclear why the observed
universe exhibits an energy density much smaller than the one
resulting from the application of quantum field theory and the
equivalence principle \cite{Weinberg}. 
Recently we pointed out that the sensitivity of present macroscopic 
small force detectors could allow, in a not too distant future, 
to experimentally verify the (passive) gravitational mass of vacuum energy, by 
detecting the extremely small forces exerted by the earth gravitational 
field on a multi-layer Casimir cavity whose vacuum energy were suitably 
modulated \cite{Callo}. These considerations 
led our search for a method to modulate the Casimir 
energy in a rigid cavity without exchanging much more energy with the system
(to avoid destroying the possibility of measuring and 
control): these conditions are
satisfied if the cavity mirrors are composed by materials that can undergo 
a superconducting transition. 
In particular we showed that the use of phase transitions offers 
also the possibility to actually measure the energy change: for a given 
temperature, the external magnetic field needed to destroy 
superconductivity, i.e. the 
critical field, is in fact proportional  to the total variation in free 
energy between the normal and superconducting state at zero field; if 
the condensation energy and Casimir variation are comparable, the 
total energy variation, and thus the critical field, of a film being part of 
a cavity can be sensibly different from that of a simple film \cite{Bimon}. The 
Aladin2 experiment has been conceived to verify this hypothesis, 
demonstrating the effect of vacuum fluctuations on a phase transition; the 
study of the possibility to modulate Casimir energy to verify its 
gravitational interaction, which was the original starting point, remains 
as a long-term motivation. The 
final measurement is foreseen for the end of 2007. Actually  the cavity is 
placed at cryogenic temperature and an external magnetic field is 
applied, parallel to the plane of the films. The applied field 
$H^{C}(T)$ necessary 
to destroy superconductivity is measured as a function 
of temperature. The expected signal is a different behavior of the 
function $H^{C}(T)$ with respect to the critical 
field $H^{F}(T)$ of a simple film. In this paper we report 
sensitivity studies on final film and cavities.

\section{Expected signal and experimental description }

The basic Casimir cavity is composed by a first thin layer (10 nm) 
of superconducting material (Al) separated by an intermediate thin 
layer (10 nm of oxide) from a third metallic layer (100 nm), not 
superconducting. The areas used are $20 \times 20 \mu m^{2}$ and 
$100 \times 100 \mu m^{2}$ to verify that the effect does not 
depend on the area.
The films and cavities are obtained by depositing the Al on the whole 
chip (some $cm^{2}$) and growing films and cavities in the same way and 
in the same time (they both have a 10 nm oxide) while the only 
difference is on the final layer, deposited only for the cavities. 
Finally, the chip is divided into samples that contain both a film and a 
cavity that thus will experience the same field, reducing the alignment 
problem to a negligible contribution. The expected signal is shown in 
Fig. 1 where the difference in transition temperature $\delta(T) 
= T_{c} - T$ is reported as a function of the  external magnetic 
field $H$. The upper curve refers to a simple film while the lower one 
to the same film being a mirror of the Casimir cavity. The theoretical 
predictions are valid from fields higher than a certain reduced critical 
field $H_{V} \approx 50$ Gauss, where the Casimir contribution to 
energy variation is simply a perturbation of condensation energy. 
Within these approximations the two curves are expected to show a constant 
difference $\Delta(T) = \delta(T)_{f} - \delta(T)_{c}$ (in our case 
$\Delta(T) \approx 0.2 mK$), not dependent on the applied field.

In the region of field lower than $H_{V}$ the Casimir contribution 
to total energy variation cannot be considered a perturbation so 
that, strictly speaking, there is not a complete theoretical 
prediction and this is reported in Fig. 1 (see Proceedings in
Nucl. Instr. Meth.) with the dotted part of the 
lower curve. The experimental apparatus is based on the commercial 
cryostat Oxford Instruments HELVLTD HelioxVL 3He, reaching the base 
temperature of 300 mK, inserted in a magnetically isolated dewar. 
The external field is generated by 
a 3.0 Gauss/mAmpere superconducting coil and the current is supplied and 
measured with a sensitivity better than 1/1000 by a multimeter HP 34401A. 
The measurement method is a standard omodine four-wire resistance. 
The actual measurement is performed by fixing the external field and storing 
$R(T)$. A set of measurements is reported in Fig. 2
(see Proceedings in Nucl. Instr. Meth.): the transition width 
is approximately 50 mK.\\

The results of sensitivity study are displayed in Fig. 1 
(see Proceedings in Nucl. Instr. Meth.) where  
$\delta(T)_{c}$ is reported as a function of the reduced field. 
It is seen that the sensitivity $\delta_{n}$ on the single measurement 
is approximatively 
$\delta_{n}$ = 0.1 mK, which can be sufficient to perform the final 
measurement.\\
It is very important to stress that the actual measurement is not 
a measurement of the absolute behavior of the field of the cavity: 
such a measurement would require an accuracy (mostly on field-cavity 
alignment) which would impose extremely difficult experimental 
constraints. Instead, our measurement will be the difference of the 
behavior of two structures (film and cavity). In Fig. 1 
(see Proceedings in Nucl. Instr. Meth.) the data are reported  
for the cavity case, while the final measurement simply would 
consist in repeating the measurement also for the film and to look for 
the eventual difference. Thus, the results reported if Fig. 1 
(see Proceedings in Nucl. Instr. Meth.) must be 
regarded as sensitivity studies: if the film (upper curve) will show 
the expected 0.2 mK shift in $\delta(T)$, our sensitivity should be 
sufficient to detect it.  

\section{AC measurements}

As stated previously there is not a complete theory describing the 
behavior of $\delta(T)_{c}$ for applied field H near (or less than) 
$H_{V}$: nonetheless this region corresponds to the physical 
region where the Casimir effect is predominant 
(i.e. it is not a perturbation of condensation 
energy), so that it is of particular interest from an experimental point 
of view. In this region, instead of studying the difference 
$\Delta(T)$ it is better to study the first derivative 
$\frac{d\delta(T)}{dH}$. 
In the case of a simple film the derivative is a straight line in the 
whole region of interest, while from general arguments \cite{Tinkam}
in the region of field $H << H_{V}$, corresponding to transition
temperatures very close to $T_{c}$, also for the cavity case the
derivative must be linearly dependent on $H$.\\
Furthermore, as is already shown in Fig. 1
(see Proceedings in Nucl. Instr. Meth.), for H sufficiently high 
the difference of shift in temperature $\Delta(T) = \delta(T)_{f} 
- \delta(T)_{c}$ is expected to be constant (independent of H), so that 
the two derivatives $\frac{d\delta(T)}{dH}_{f,c}$ should converge to the 
same value. (We recall that the actual value of 
$\Delta(T) = 0.2 mK $ in our case).

A possible situation is reported in Fig. 3 (see Proceedings in 
Nucl. Instr. Meth.). Whatever will be the 
actual curve, the relative difference of the two cases is expected 
to be quite high, of the order of some tens percent. Indeed, this 
corresponds to the fact that the difference 
$\Delta(T) = 0.2 mK $ is expected to be reached already for 
$\frac{H}{H_{V}} \approx 3$ (where the shift in Casimir energy is a 
tenth of the condensation energy) where $\delta(T)_{f,c} \approx 
0.6(0.4) mK$, respectively. From an experimental point of view this 
is very encouraging because, as shown in Fig. 3
(see Proceedings in Nucl. Instr. Meth.), the sensitivity 
is far better than few percent on each point, so that we expect 
to have the possibility to detect the vacuum fluctuation 
effect with a good signal-to-noise ratio.

\end{document}